\documentclass[11pt]{article}

\usepackage{graphicx}
\usepackage{color}
\usepackage{hyperref}
\usepackage{setspace}
\usepackage{a4wide}
\usepackage{latexsym}

\title{A Stochastic Model for the Evolution of the Web}

\author{Mark Levene, Trevor Fenner, George Loizou and Richard Wheeldon \\
School of Computer Science and Information Systems \\
Birkbeck College, University of London \\
London WC1E 7HX, U.K. \\ \{mark,trevor,george,richard\}@dcs.bbk.ac.uk}

\date{}

\begin{document}

\maketitle

\newtheorem{theorem}{Theorem}[section]
\newtheorem{corollary}[theorem]{Corollary}
\newtheorem{lemma}[theorem]{Lemma}
\newtheorem{proposition}[theorem]{Proposition}
\newtheorem{definition}{Definition}[section]
\newtheorem{algorithm}{Algorithm}
\newtheorem{example}{Example}[section]

\begin{abstract}

Recently several authors have proposed stochastic models of the
growth of the Web graph that give rise to power-law distributions.
These models are based on the notion of preferential attachment
leading to the ``rich get richer'' phenomenon. However, these models
fail to explain several distributions arising from empirical results,
due to the fact that the predicted exponent is not consistent with
the data. To address this problem, we extend the evolutionary model
of the Web graph by including a non-preferential component, and we
view the stochastic process in terms of an urn transfer model. By
making this extension, we can now explain a wider variety of
empirically discovered power-law distributions provided the exponent
is greater than two. These include: the distribution of incoming
links, the distribution of outgoing links, the distribution of pages
in a Web site and the distribution of visitors to a Web site. A
by-product of our results is a formal proof of the convergence of the
standard stochastic model (first proposed by Simon).

\end{abstract}

\section{Introduction}

A power-law distribution is a function of the form
\begin{displaymath}
f(i) = C \ i^{- \tau},
\end{displaymath}
where $C$ and $\tau$ are positive constants. Power-law distributions
are {\em scale-free} in the sense that if $i$ is rescaled by
multiplying it by a constant, then $f(i)$ would still be proportional
to $i^{- \tau}$.

\smallskip

Power-law distributions are abundant, for example
Zipf's law \cite{RAPO82}, which states that relative frequency of words in a
text is inversely proportional to their rank, and
Lotka's law \cite{NICH89}, which is an inverse square law stating that
the number of authors making $n$ contributions is proportional to $n^{-2}$.
(We refer the reader to \cite{SCHR91} for more examples of power-law
distributions.)

\medskip

Recently several researchers have detected power-law distributions in
the Internet \cite{FALO99} and World-Wide-Web \cite{BROD00,DILL01}
topologies. In order to understand how these power-law distributions
emerge and how the Web has evolved and is evolving, several
researchers have recently been studying stochastic models of graphs
which give rise to such distributions. One particular power-law
phenomenon that has attracted attention is the distribution of
incoming links to a Web page. This distribution is important, since a
link from Web page $P$ to Web page $Q$ can be viewed as a
recommendation of page $Q$; thus Web pages having more incoming links
are more highly recommended and therefore potentially of higher
quality. This observation is the basis of Google's PageRank algorithm
\cite{HENZ01}.

\smallskip

Albert et al. \cite{ALBE00b} studied a stochastic model of growth and
{\em preferential attachment}, where new links to existing Web pages
are added in proportion to the number of incoming links these Web
pages already have. Their theoretical model predicts an exponent
$\tau = 3$, which is not in agreement with the value of approximately
$2.1$ obtained from the study reported in \cite{BROD00}. Dorogovtsev
et al. \cite{DORO00a} generalise Albert et al.'s model and predict an
exponent greater than two. More precisely, they obtain the value $2 +
A/m$ for the exponent, where $A$ is the initial attractiveness of a
newly created Web page and $m$ is the number of new links added to
the Web graph at each step of the stochastic process. This exponent
value is consistent with the empirical value of the exponent of the
distribution of incoming links provided $A/m$ is sufficiently small.
Bornholdt and Ebel \cite{BORN00} pointed out that the stochastic
process proposed by Simon \cite{SIMO55} in 1955 can also offer an
explanation of the power-law distribution. (We note that during the
period of 1959-1961 there was a fierce debate between Mandelbrot and
Simon in {\em Information and Control} on the validity of Simon's
model \cite{MAND59}.) In reply to Bornholdt and Ebel, Dorogovtsev et
al. \cite{DORO00b} note that the model they describe in
\cite{DORO00a} essentially coincides with Simon's model.

\medskip

The models discussed above are based on the process of preferential
attachment and do not take into account the fact that links may also
be added or removed randomly through a {\em non-preferential
process}. By this we mean that the probability of adding or removing
a link to a particular Web page may be influenced by factors other
than the popularity of that Web page, where popularity is measured by
the number of incoming links. Our main contribution in this paper is
to extend Simon's model \cite{SIMO55} with a non-preferential
component and view the stochastic process in terms of an urn transfer
model \cite{JOHN77}. (We note that, at the end of Section 3 of his
seminal paper, Simon suggested adopting a mixture of preferential and
non-preferential components but did not develop the idea.) By making
this extension we can explain a wider variety of empirically
discovered power-law distributions than can be explained with Simon's
original model. These include: the distribution of incoming links,
the distribution of outgoing links, the distribution of pages in a
Web site and the distribution of visitors to a Web site.

\medskip

The rest of the paper is organised as follows. In
Section~\ref{sec:urn} we present an urn transfer model that
generalises Simon's original model. In Section~\ref{sec:model} we
demonstrate how this can provide a stochastic model for the evolution
of the Web that is consistent with a wide range of empirical data.
Finally, in Section~\ref{sec:concluding} we give our concluding
remarks. The proofs of some of the mathematical results are given in
the Appendix. As far as we are aware, our convergence proof given in
the Appendix is the first formal proof validating Simon's model $-$
it does not rely on the mean-field theory approach, as for example in
\cite{BARA99}.

\section{An Urn Transfer Model}
\label{sec:urn}

We now present an {\em urn transfer model} \cite{JOHN77} for a
stochastic process that we will use in Section~\ref{sec:model} to
analyse the evolution of the Web graph. Our model is an extension of
Simon's stochastic process \cite{SIMO55}, which was originally
described in terms of the underlying process leading to the
distribution of words in a piece of text. Simon's stochastic process
is essentially a birth process, where there is a constant probability
$p$ that the next word is a new word and, given that the next word
has already appeared, its probability of occurrence is proportional
to the previous number of occurrences of that word. We extend Simon's
model by setting the probability of occurrence of a word, given that
it has already appeared, to be a weighted average of the preferential
probability, as described above, and the uniform probability if all
words were equiprobable. As we noted in the introduction, this
extension was already proposed by Simon at the end of Section 3 of
his paper. Simon set out this extension in equation (3.7) and
presented a tentative solution in equation (3.8). As we will see in
Section~\ref{sec:model}, our extension of Simon's model makes sense
in the context of the Web, since, for example, links to a Web site
are often added or removed in a random fashion without taking into
consideration the ``attractiveness'' of the site in terms of how many
links it already has. In our urn model, urns contain balls which have
pins attached to them. For example, balls could represent Web pages
and pins could represent inlinks or outlinks. An urn would then
correspond to the set of Web pages having a specific number of links.
We now describe our urn model in detail.

\medskip

\renewcommand{\thefootnote}{\fnsymbol{footnote}}

We assume a countable number of urns, $urn_i$, $i = 1,2,3, \ldots$,
where each ball in $urn_i$ has $i$ pins attached to it. Initially, at
stage $k =1$ of the stochastic process, all the urns are empty except
$urn_1$ which has one ball in it. Let $F_i(k)$ be the number of balls
in $urn_i$ after $k$ steps of the stochastic process, so $F_1(1) =
1$, and let $p$ and $\alpha$ be parameters, with $0 < p < 1$ and
$\alpha > -1$.\footnote[1]{The reader should note that in
\cite{SIMO55} the quantity $\alpha$ corresponds to our parameter
$p$.} Then, at stage $k+1$ of the stochastic process for $k \ge 1$,
one of two things may occur:

\renewcommand{\labelenumi}{(\roman{enumi})}
\begin{enumerate}
\item with probability $p_{k+1}$, where
\begin{equation}\label{eq:prob2}
p_{k+1} = 1 - \frac{(1-p) \sum_{i=1}^k (i + \alpha) F_i(k)}{k (1 +
\alpha p) + \alpha (1-p)},
\end{equation}
a new ball is added to $urn_1$ (provided that $0 \le p_{k+1} \le 1$)
or,

\item with probability $1 - p_{k+1}$, an urn is selected $-$ $urn_i$
being chosen with probability
\begin{equation}\label{eq:prob1}
\frac{(1-p) (i + \alpha) F_i(k)}{k (1 + \alpha p) + \alpha (1-p)},
\end{equation}
for $1 \le i \le k$; then one ball from $urn_i$ is transferred to
$urn_{i+1}$. This is equivalent to attaching an additional pin to the
ball chosen from $urn_i$ and moving it to its ``correct'' urn. The
probability (\ref{eq:prob1}) is a combination of a preferential
component (proportional to the number of pins in $urn_i$) and a
non-preferential component (proportional to the number of balls in
$urn_i$). (We note that the denominator appearing in (\ref{eq:prob2})
and (\ref{eq:prob1}) has been chosen so that $p_{k+1}$, the expected
value of the probability of adding a ball, is $p$; see
(\ref{eq:expected2}) below.)
\end{enumerate}
\smallskip

At each stage we either add a new ball with one pin or add a pin to
an existing ball and move the ball to the next urn up, so at stage
$k$ the total number of pins is $k$, i.e.
\begin{displaymath}
\sum_{i=1}^k i F_i(k) = k.
\end{displaymath}
\smallskip

It is obvious that $F_i(k) = 0$ for any $i > k$. We call the above
model the {\em $p_k$-model}.

\medskip

Let $B(k) = \sum_{i} F_i(k)$, the total number of balls in all the
urns. We can now simplify (\ref{eq:prob2}) to
\begin{equation}\label{eq:simplify}
p_{k+1} = 1 - \frac{(1-p) (k + \alpha B(k))}{k (1 + \alpha p) +
\alpha (1-p)}.
\end{equation}
\medskip

Since it is clear from (\ref{eq:simplify}) that $p_{k+1} < 1$, in
order for $p_{k+1}$ to be well defined, we must have $p_{k+1} \ge 0$
for $k \ge 1$, i.e.
\begin{equation}\label{eq:prob3}
(1-p) (k + \alpha B(k)) \le k (1 + \alpha p) + \alpha (1-p).
\end{equation}
\smallskip

In the Appendix we show that $p_{k+1}$ is always well defined (i.e.
non-negative) for all $k$ when $p \ge 1/2$, but only if
\begin{equation}\label{eq:illdefined}
\alpha \le \frac{p}{1 - 2 p}
\end{equation}
when $p < 1/2$. In the discussion at the end of
Section~\ref{sec:model} we suggest that, in practice, starting from a
typical initial configuration of balls in the urns, it is likely that
$p_{k+1}$ will be well defined for all $k$, even if
(\ref{eq:illdefined}) does not hold.

\medskip

We next make a small digression to explain our use of
(\ref{eq:prob1}) rather than the more natural definition of the
probability as
\begin{equation}\label{eq:natural}
\frac{(1-p) (i + \alpha) F_i(k)}{\sum_{i=1}^k (i + \alpha) F_i(k)} =
\frac{(1-p) (i + \alpha) F_i(k)}{k + \alpha B(k)}.
\end{equation}
\medskip

In order to find a solution for the expected value of $F_i(k)$, we
would need to take the expected value of (\ref{eq:natural}); this is
problematic since the random variables $B(k)$ and $F_i(k)$ are {\em
not} independent and it is therefore not clear how to calculate the
expectation of the right-hand expression in (\ref{eq:natural}). We
observe that this problem does not arise in Simon's original model
\cite{SIMO55}, since in this case we have $\alpha = 0$ and the
denominator reduces to the constant $k$ in both (\ref{eq:prob1}) and
(\ref{eq:natural}). In our case, when $\alpha$ is not necessarily
zero, by using (\ref{eq:prob1}) instead of the more natural
(\ref{eq:natural}), there is no problem in computing the expectation
of $F_i(k)$, since the parameter $p$ is a constant, allowing us to
find the expected value of $F_i(k)$ by using the linearity of
expectations.

\medskip

In the Appendix we prove the following results for the expectations
of $B(k)$ and $p_k$ for $k > 1$, namely
\begin{equation}\label{eq:expected1}
E(B(k)) = E \bigg( \sum_{i=1}^k F_i(k) \bigg) = 1 + (k-1) p
\end{equation}
and
\begin{equation}\label{eq:expected2}
E(p_k) = p.
\end{equation}
(We note that $E(B(1)) = B(1) = 1$.)

\medskip

Thus, in terms of expectations (i.e. using a mean-field theory
approach), it is possible to describe the urn transfer model as a
``more natural'' stochastic process, where at each stage $k$, for $k
> 1$, either
\begin{enumerate}
\item a new ball is inserted into $urn_1$ with probability $p$, or

\item with probability $1 - p$ an urn is chosen, the probability of choosing
$urn_i$ being proportional to $(i + \alpha) F_i(k)$, and then one
ball from $urn_i$ is transferred to $urn_{i+1}$.
\end{enumerate}
\smallskip

We stress that, since this model uses the expectations of the random
variables $p_{k+1}$ rather than the random variables themselves, it
is only an approximation of our urn transfer model. This model, which
we call the {\em $p$-model}, is in fact the ``more natural'' model
discussed above that uses (\ref{eq:natural}) instead of
(\ref{eq:prob1}).

\medskip

We note that we could modify the initial condition of our stochastic
process so that, for example, $urn_1$ would initially contain $\delta
> 1$ balls instead of one, or more generally that a finite number of
urns would initially be non-empty with some prescribed number of
balls in each. As can be seen from the development of the model
below, as $k$ tends to infinity, such a change in the initial
conditions will not have an effect on the asymptotic distribution of
the balls in the urns.

\medskip

We call the transfer of a ball as a result of (ii) above a mixture of
{\em preferential} and {\em non-preferential} transfer. When $\alpha
= 0$, then the transfer is purely preferential otherwise
non-preferential transfer takes a part in the process.

\medskip

Following Simon \cite{SIMO55}, we now state the equations for the
$p_k$-model. For $i > 1$ (including $i > k$), we have:
\begin{equation}\label{eq:1st-ss0}
E_k (F_i(k+1)) = F_i(k) + \beta_k \bigg((i-1+ \alpha) F_{i-1}(k) -
(i+ \alpha) F_i(k) \bigg),
\end{equation}
\smallskip
where $E_k (F_i(k+1))$ is the expected value of $F_i(k+1)$ given the
state of the model at stage $k$, and
\begin{displaymath}
\beta_k = \frac{1 - p}{k (1 + \alpha p) + \alpha (1-p)},
\end{displaymath}
the normalising constant used in (\ref{eq:prob1}).

\smallskip

Equation (\ref{eq:1st-ss0}) gives the expected number of balls in
$urn_i$ as the previous number of balls in that urn plus the
difference between the probability of increasing the number of balls
in $urn_i$, which is equal to the probability of choosing $urn_{i-1}$
in step (ii) of our urn transfer model, and the probability of
decreasing the number of balls in $urn_i$, which is equal to the
probability of choosing $urn_i$.

\medskip

In the boundary case, $i = 1$, we have
\begin{equation}\label{eq:1st-initial}
E_k (F_1(k+1)) = F_1(k) + p_{k+1} - \beta_k (1 + \alpha) F_1(k),
\end{equation}
\smallskip
which describes the expected number of balls in $urn_1$, which is the
previous number of balls in the first urn plus the difference between
the probability of inserting a new ball in $urn_1$ and the
probability of transferring a ball from $urn_1$ to $urn_2$.

\medskip

Now letting
\begin{displaymath}
\beta = \frac{1 - p}{1 + \alpha p},
\end{displaymath}
we see that $k \beta_k \approx \beta$ for large $k$. In fact, for $k
\ge 1$,
\begin{equation}\label{eqZ}
\beta - k \beta_k = \alpha \beta \beta_k.
\end{equation}
\smallskip

Using the facts that $0 < p < 1$ and $\alpha > - 1$, it is also easy
to see that
\begin{equation}\label{eqY1}
0 < \beta < 1,
\end{equation}
and, for $k \ge 1$,
\begin{equation}\label{eqY2}
0 < \beta_k < \frac{1}{k + \alpha}.
\end{equation}
\medskip

Since the right-hand sides of (\ref{eq:1st-ss0}) and
(\ref{eq:1st-initial}) are linear in the random variables, using
(\ref{eq:expected2}), we may take expectations to obtain
\begin{equation}\label{eq:ss0}
E(F_i(k+1)) = E(F_i(k)) + \beta_k \bigg((i-1+ \alpha) E(F_{i-1}(k)) -
(i+ \alpha) E(F_i(k)) \bigg)
\end{equation}
for $i > 1$, and
\begin{equation}\label{eq:initial}
E(F_1(k+1)) = E(F_1(k)) + p - \beta_k (1 + \alpha) E(F_1(k)).
\end{equation}
\medskip

In order to solve equations (\ref{eq:ss0}) and (\ref{eq:initial}) we
show that $E(F_i (k)) / k$ tends to a limit $f_i$, as $k$ tends to
infinity. Assume for the moment that this is the case, then, letting
$k$ tend to infinity, $E(F_i(k+1)) - E(F_i(k))$ tends to $f_i$ and
$\beta_k E(F_i(k))$ tends to $\beta f_i$; so (\ref{eq:ss0}) and
(\ref{eq:initial}) yield
\begin{equation}\label{eq:ss2}
f_i = \beta \bigg( (i - 1 + \alpha) f_{i-1} - (i + \alpha) f_i \bigg)
\end{equation}
for $i > 1$, and
\begin{equation}\label{eq:ss1}
f_1 = p - \beta (1 + \alpha) f_1.
\end{equation}
\medskip

Now let us {\em define} $f_i$, $i \ge 1$, by the recurrence relation
(\ref{eq:ss2}) with boundary condition (\ref{eq:ss1}). We may then
let
\begin{equation}\label{eq:epsilon}
E(F_i(k)) = k (f_i + \epsilon_{i,k}),
\end{equation}
and in the Appendix we prove that $\epsilon_{i,k}$ tends to zero as
$k$ tends to infinity. This justifies our assumption that $E(F_i(k))
/ k$ tends to $f_i$ as $k$ tends to infinity. We therefore see that
$f_i$ is the asymptotic expected rate of increase of the number of
balls in $urn_i$.

\medskip

From (\ref{eq:ss2}) and (\ref{eq:ss1}) we obtain
\begin{equation}\label{eq:ss4}
f_i = \frac{\beta (i - 1 + \alpha)}{1 + \beta (i + \alpha)} \ f_{i-1}
\end{equation}
and
\begin{equation}\label{eq:ss3}
f_1 = \frac{p}{1 + \beta (1 + \alpha)},
\end{equation}
respectively.

\smallskip

Now, on repeatedly using (\ref{eq:ss4}), we get
\begin{eqnarray}\label{eq:gamma}
f_i & = & \frac{\rho \ p \ (1 + \alpha) \ (2 + \alpha) \ \cdots \ (i
- 1 + \alpha)}{(1 + \rho + \alpha) \ (2 + \rho + \alpha) \
\cdots \ (i + \rho + \alpha)} \nonumber \\
& = & \frac{\rho \ p \ \Gamma(i + \alpha) \ \Gamma(1 + \rho +
\alpha)} {\Gamma(1 + \alpha) \ \Gamma(i + 1 + \rho + \alpha)},
\end{eqnarray}
where $\rho = 1 / \beta$ and $\Gamma$ is the gamma function
\cite[6.1]{ABRA72}.

\medskip

It follows that for large $i$,
on using Stirling's approximation \cite[6.1.39]{ABRA72}, we have
\begin{equation}\label{eq:ss5}
f_i \sim C \ i^{- (1 + \rho)},
\end{equation}
where $C$ is independent of $i$ and $\sim$ means {\em is asymptotic
to}. Thus we have derived in (\ref{eq:ss5}) a general power-law
distribution for $f_i$, with exponent $1 + \rho$. An obvious
consequence of (\ref{eq:ss4}) is that $f_i > f_{i+1}$, i.e.
asymptotically there are more balls in $urn_i$ than in $urn_{i+1}$.

\smallskip

It follows from (\ref{eq:natural}), (\ref{eq:expected1}) and
(\ref{eq:expected2}) that (\ref{eq:ss2}) and (\ref{eq:ss1}) will also
hold for the asymptotic distribution for the $p$-model obtained using
the mean-field theory approach. Hence, on the assumption that this
approach is valid, the asymptotic distribution will be the same as
for the $p_k$-model, as given by (\ref{eq:gamma}) and (\ref{eq:ss5}).

\medskip

When $\alpha = 0$ then the extended model reduces to Simon's original
model and by increasing $\alpha$ the exponent will increase
accordingly. In any case the exponent is always greater than $2$, so
the expected number of pins per ball is finite. The constraint that
$\rho > 1$ is equivalent to the condition that $\alpha > -1$. Another
way to understand this constraint is that if $\alpha \le -1$ then the
first urn will never be chosen in case (ii) of the stochastic
process, and thus no ball will ever be transferred out of $urn_1$.
When $\rho$ is close to $1$ we obtain Lotka's law \cite{NICH89},
which is an inverse square power-law; see also Price's cumulative
advantage distribution leading to Lotka's law \cite{PRIC76,KOEN95}.

\smallskip

In many real situations, such as the Web, $p$ is generally small. For
example, if we interpret balls as Web pages and the number of pins
attached to a ball as the number of links incoming to that Web page,
then we expect the ratio of pages to links to be quite small, say
0.1, and thus the exponent of the power-law to be just over two. If
the value of $p$ and the power-law exponent are obtained from
empirical evidence, we may find a discrepancy from Simon's original
model, i.e. when $\alpha = 0$. Our current extension can explain this
discrepancy through the non-preferential component as long as the
exponent is greater than two.

\section{A Model for the Evolution of the Web}
\label{sec:model}

We now describe a discrete stochastic process by which the Web graph
could evolve. At each time step the state of the Web graph is a
directed graph $G = (N, E)$, where $N$ is its node set and $E$ is its
link set. In this case $F_i(k)$, $i \ge 1$, is the number of nodes in
the Web graph having $i$ incoming links; $F_i(k)$ induces an
equivalence class of nodes in $N$ all having $i$ incoming links. We
note that although we have chosen $i$ to denote the number of incoming
links, $i$ could alternatively denote the number of outgoing links, the
number of pages in a Web site or any other reasonable parameter.

\smallskip

Consider the evolution process of the Web graph with respect to the
number of nodes having $i$ incoming links at the $k$th step of the
process. Initially $G$ contains just a single node. At each step one
of two things can happen. With probability $p$ a new node is added to
$G$ having one incoming link. In the $p$-model, this is equivalent to
placing a new ball in $urn_1$. With probability $1-p$ a node is
chosen to receive a new incoming link, with the probability of
choosing a given node being proportional to $(i + \alpha)$, where $i$
is the number of incoming links the node currently has. In the
$p$-model this is equivalent to a mixture of preferential and
non-preferential transfer of a ball from $urn_i$ to $urn_{i+1}$; the
mixture level depends on the value of the parameter $\alpha$.

\smallskip

When $\alpha = 0$ a node is chosen according to preferential
attachment, i.e. in proportion to the number of inlinks to that node.
In this case the number of inlinks to a Web page could be interpreted
as a measure of how important or recommended the Web page is. A
situation when $\alpha > 0$ might occur if there is a choice of Web
pages to link to and the actual decision of which links are put in
place has a random component. For example, if we were to add several
links to Web pages pertaining to Zipf's law to our Web site, we might
randomly choose them from a resource containing hundreds of such
links. A situation when $\alpha < 0$ might occur if we consider
internal links within Web pages not to be relevant when measuring the
distribution of inlinks to Web pages. The justification for this view
is that internal links do not contribute to the external
``visibility'' of a Web page.

\medskip

We now look at some of the measurements of the Web graph which were
reported recently. Broder et al. \cite{BROD00} reported a power-law
distribution with exponent $2.09$ for the number of incoming links
(referred to as {\em inlinks}) to a node. This value was derived from
a 203 million node crawl of the Web graph. The average number of
inlinks per Web page was measured at about 8 \cite{KUMA99b}, which
gives us a value of $0.125$ for $p$. We can compute $\alpha$ by
\begin{displaymath}
\alpha = \frac{\rho (1 - p) - 1}{p}.
\end{displaymath}
\medskip

Thus a more accurate model of the stochastic process generating the
distribution of incoming links would assume $\alpha \approx - 0.37$
rather than $\alpha = 0$. (It would not be unreasonable in this case
to assume Simon's model, i.e. $\alpha = 0$, which would give an
exponent of $2.14$, since the small difference in the exponents may
be due to statistical error.)

\smallskip

Looking at the outgoing links (referred to as {\em outlinks}) from a
node, Broder et al. \cite{BROD00} reported a power-law distribution
with exponent $2.72$. Moreover, the average number of outlinks per
Web page was measured at about $7.2$ \cite{KUMA00a}, which gives a
value of $0.14$ for $p$. Thus to get an exponent of $2.72$ we would
have to assume $\alpha \approx 3.42$. However, Simon's original model
would predict an exponent of about $2.16$ for outlinks, similar to
that for inlinks. The positive value of $\alpha$ may have occurred
due to the fact that outlinks are often created for reasons other
than preferential attachment, for example, in order to maintain the
local structure of a Web site.

\smallskip

Another interpretation of $i$ is the number of pages within Web sites
(referred to as {\em webpages}). In this case, Huberman and Adamic
\cite{HUBE99} reported a power-law distribution with exponent $1.85$,
derived from a 250,000 Web site crawl. Our model cannot explain this
observation as the exponent is less than two. A more recent result
from a private communication with Adamic reported an exponent of 2.2,
derived from a 1.6 million Web site crawl; the difference is possibly
due to a different crawling strategy. To calculate $p$ we can
estimate the size of the Web to be $2.1$ billion pages \cite{MURR00}
distributed over approximately $113.5$ million Web sites (this
number, which was reported on
\href{http://www.netsizer.com}{www.netsizer.com} during the first
quarter of 2001, refers to the number of Internet hosts, so it is an
over-estimate of the number of Web sites). Thus we can derive a value
$0.054$ for $p$; in reality $p$ will be even closer to zero. To get
an exponent of $2.2$ we would have to assume $\alpha \approx 2.50$.
This gives a more accurate description than we would obtain from
Simon's original model, which would predict an exponent of $2.06$.
The positive value of $\alpha$ may have occurred due to the fact that
pages in Web site may be created in different ways, for example,
pages may be created dynamically by a content management system. This
may tend to increase the number of pages by adding certain generated
pages.

\smallskip

As a final interpretation, let $i$ be the number of users visiting a
Web site during the course of a day (referred to as {\em visitors}).
In this case, Adamic and Huberman \cite{ADAM00} reported a power-law
distribution with exponent $2.07$, derived from access logs of 60,000
AOL users accessing 120,000 Web sites. Now, from
\href{http://www.netsizer.com}{www.netsizer.com} we obtain the
statistic that in the USA there were $72.7$ million Internet hosts
and $166.6$ million users at the beginning of 2001. Moreover, from
\href{http://www.netvalue.com}{www.netvalue.com} we obtain the
statistic that, on average, users visit about $1.93$ different Web
sites per day. So, we derive the value $72.7 / (166.6
* 1.93) \approx 0.226$ for $p$, on the assumption that each Web site gets
visited at least once per day. Thus to get an exponent of $2.07$ we
would have to assume $\alpha \approx - 0.76$. However, Simon's
original model would predict an exponent of about $2.29$. The
negative value of $\alpha$ may have occurred due to the fact that
some visitors to a Web site may tend to avoid well-trodden sites
which may have too much commercial content.

\medskip

In order to validate our model, we programmed a simulation of the
stochastic model using the parameter values we have derived for $p$
and $\alpha$ and compared the exponent values obtained with the
reported empirical values. (Our simulation is in the spirit of Simon
and Van Wormer's Monte Carlo simulation, whose intention was to test
how good the estimates of the original model are \cite{SIMO63}.) We
repeated the simulation five times using the $p_k$-model, and five
times using the $p$-model. Each simulation was carried out for
200,000 iterations, and for the purpose of regression we considered
only the first 25 urns. The results of our simulations are presented
in Table~\ref{table:simulations}; in all cases the correlation
coefficient of the regression analysis was close to one. The
discrepancy between the simulated values and the empirical values can
be attributed in part to the fact that (\ref{eq:ss5}) is only an
asymptotic approximation to (\ref{eq:gamma}). It is also possible
that running the simulations for a much larger number of iterations
would give more accurate results.

\begin{table}[ht]
\begin{center}
\begin{tabular}{|l|c|c|c|} \hline
Interpretation & Empirical & $p_k$-model & $p$-model \\ \hline
inlinks & 2.09 & 2.096 & 2.094 \\
outlinks & 2.72 & 2.714 & 2.675 \\
webpages & 2.2 & 2.122 & 2.208 \\
visitors & 2.07 & 2.131 & 2.179 \\ \hline
\end{tabular}
\end{center}
\caption{\label{table:simulations} Power law exponents of simulation
results}
\end{table}
\medskip

For outlinks and webpages we restarted the $p_k$-model simulation
whenever the computed value of $p_{k+1}$ was ill-defined, i.e.
negative; only a moderate number of restarts were necessary. From
(\ref{eq:simplify}) it can be shown that, for $k > 1$, $\mid \!
p_{k+1} - p_k\! \mid$ is of the order of $1 / k$. This indicates that
for large $k$ it is very unlikely that $p_{k+1}$ will be ill-defined,
given that $p_j$ is well defined for $j \le k$. In practice, if
instead of starting with just one ball in $urn_1$, we start from a
typical initial configuration with a modest number of balls in the
urns, it is likely that $p_{k+1}$ will be well defined for all $k$.

\begin{table}[ht]
\begin{center}
\begin{tabular}{|c|c|c|c|c|} \hline
Batch & Overall & $k \le 10$ & Average $k$ & Max $k$\\
\hline
1 & 66 & 89\% & 3.78 & 21 \\
2 & 63 & 90\% & 3.86 & 26 \\
3 & 63 & 90\% & 3.34 & 13 \\
4 & 60 & 88\% & 3.68 & 30 \\
5 & 63 & 90\% & 3.73 & 22 \\
6 & 65 & 92\% & 3.49 & 22 \\
7 & 61 & 92\% & 3.50 & 17 \\
8 & 64 & 94\% & 3.54 & 22 \\
9 & 64 & 86\% & 4.19 & 34 \\
10 & 59 & 92\% & 3.49 & 21 \\
\hline Average & 63 & 90\% & 3.66 & 23 \\ \hline
\end{tabular}
\end{center}
\caption{\label{table:restarts} Statistics for restarts, with $p =
0.15$ and $\alpha = 3.5$}
\end{table}
\medskip

To illustrate this point, let us now examine more closely the
situation regarding restarts for outlinks, rounding off $p$ to be
$0.15$ and $\alpha$ to be $3.5$. It can be verified that the
probability that $p_3$ be ill-defined is $0.15$, that $p_4$ be
ill-defined is about $0.1905$, that $p_5$ be ill-defined is about
$0.1769$ and that $p_6$ be ill-defined is $0$. Thus the total
probability of $p_k$ being ill-defined for $k \le 6$ is about
$0.5174$. Table~\ref{table:restarts} shows the values of a simulation
where the $p_k$-model was run $1000$ times in batches of $100$ runs
each. Whenever $p_k$ was ill-defined for a given run, this run was
considered to be a restart and the simulation moved on to the next
run. The second column shows the numbers of restarts within the
batch, the third column shows the percentage of the restarts observed
with $p_k$ ill-defined for $k \le 10$, the fourth column shows the
average stage at which the restarts became ill-defined and the fifth
column shows the maximum stage at which any restart became
ill-defined. Thus, if the process is well defined for, say, $50$ or
more stages, it is very unlikely to become ill-defined at a later
stage.

\section{Concluding Remarks}
\label{sec:concluding}

We have extended Simon's classical stochastic model by adding to it a
non-preferential component which is combined with preferential
attachment. When viewing this stochastic process in terms of an urn
transfer model, this amounts to choosing balls proportional to the
number of times they have previously been selected (i.e. the number
of pins) plus a constant $\alpha > - 1$. From the equations of this
process we derived an asymptotic formula for the exponent of the
resulting power-law distribution. As far as we are aware our proof
given in the Appendix is the first formal proof of the convergence of
Simon's model; unlike in previous work, we do not rely on the
mean-field theory approach.

\smallskip

Utilising our result we are able to explain several power-law
distributions in the Web graph, which we now summarise. For the
distribution of incoming links we derived $\alpha \approx - 0.37$,
for the distribution of outgoing links we derived $\alpha \approx
3.42$, for the distribution of pages in a Web site we derived $\alpha
\approx 2.50$ and, finally, for the distribution of visitors to a Web
site we derived $\alpha \approx - 0.76$. In all cases our extension
of Simon's original model can better explain the exponent of the
power-law distribution, indicating that there is some mixture of
preferential and non-preferential attachment in the selection
process.

\medskip

The power law distribution that we have established can be stated as
a hypothesis: {\em in order to explain the evolution of the Web graph
both preferential and non-preferential processes are at work}. This
hypothesis is more consistent with empirical data than the one which
utilises only preferential attachment. Our model is still limited to
the cases where the exponent of the power-law distribution is greater
than two. We are currently investigating a possible model which could
yield an exponent less than two.

\appendix
\section{Appendix : Proofs}
\label{sec:proofs}

We first prove (\ref{eq:expected1}) and (\ref{eq:expected2}). Since
at stage $k+1$ we add a new ball with probability $p_{k+1}$,
\begin{displaymath}
E_k(B(k+1)) = B(k) + p_{k+1},
\end{displaymath}
so, taking expectations,
\begin{equation}\label{eq:zero}
E(B(k+1)) = E(B(k)) + E(p_{k+1}).
\end{equation}
\smallskip

\begin{lemma}\label{lemma:expectations}
\begin{rm}
For $k > 1$,
\begin{equation}\label{eq:lemma-expected1}
E(B(k)) = E \bigg( \sum_{i=1}^k F_i(k) \bigg) = 1 + (k-1) p
\end{equation}
and
\begin{equation}\label{eq:lemma-expected2}
E(p_k) = p.
\end{equation}
\end{rm}
\end{lemma}
{\em Proof.} We prove the result by induction on $k$. For $k = 2$,
remembering that $B(1) = 1$ and using (\ref{eq:simplify}), it is easy
to see that $p_2 = p$, and thus, by using (\ref{eq:zero}), that
\begin{displaymath}
E(B(2)) = 1 + E(p_2) = 1 + p.
\end{displaymath}
\smallskip

Now assume that (\ref{eq:lemma-expected1}) and
(\ref{eq:lemma-expected2}) hold for some $k$, $k > 1$. Then,
\begin{displaymath}
E(p_{k+1}) = 1 - \frac{(1-p)(k + \alpha E(B(k)))}{k (1 + \alpha p) +
\alpha (1-p)} = 1 - \frac{(1-p)(k + \alpha (1 + (k-1)p))}{k (1 +
\alpha p) + \alpha (1-p)} = p
\end{displaymath}
and thus, using (\ref{eq:zero}),
\begin{displaymath}
E(B(k+1)) = 1 + (k-1) p + p = 1 + k p. \quad \Box
\end{displaymath}
\medskip

We now consider condition (\ref{eq:prob3}) needed for $p_{k+1}$ to be
well defined.

\begin{lemma}\label{lemma:welldefined}
\begin{rm}
$p_{k+1}$ is always well defined (i.e. non-negative) for all $k$ when
$p \ge 1/2$, but only if
\begin{displaymath}
\alpha \le \frac{p}{1 - 2 p}
\end{displaymath}
when $p < 1/2$.
\end{rm}
\end{lemma}
{\em Proof.} In order that $p_{k+1} \ge 0$, condition
(\ref{eq:prob3}) must hold. This is equivalent to
\begin{equation}\label{eq:bk}
\alpha (B(k) - 1) \le \frac{k p \ (1 + \alpha)}{1-p}.
\end{equation}
\smallskip

There are three cases to consider:
\renewcommand{\labelenumi}{(\Roman{enumi})}
\begin{enumerate}
\item When $\alpha = 0$, there are no restrictions on $p$.

\item When $-1 < \alpha < 0$, it is straightforward to see that again
there are no restrictions on $p$ since, in this case, the maximum
value of the left-hand side of (\ref{eq:bk}) is zero, when $B(k) =
1$.

\item When $\alpha
> 0$, we see from (\ref{eq:bk}) that we must have
\end{enumerate}
\begin{displaymath}
p \ge \frac{\alpha (B(k)-1)}{\alpha(B(k) - 1) + k (1 + \alpha)}.
\end{displaymath}
\smallskip

Setting $B(k)$ to its maximum value $k$, this requires that
\begin{equation}\label{eq:pkmax}
p \ge \frac{\alpha (k-1)}{\alpha (2 k - 1) + k},
\end{equation}
\smallskip
which will hold for all $k$ provided
\begin{displaymath}
p \ge \frac{\alpha}{2 \alpha + 1},
\end{displaymath}
in particular this holds for all $\alpha$ when $p \ge 1/2$ . However,
for $p < 1/2$, for (\ref{eq:pkmax}) to hold for all $k$ we need
\begin{displaymath}
\alpha \le \frac{p}{1 - 2 p}. \quad \Box
\end{displaymath}
\medskip


We conclude by proving that as $k$ tends to infinity $E(F_i(k)) / k$
tends to $f_i$, justifying our derivation of the asymptotic
distribution of the balls in the urns. We first state some useful
properties of $f_i$, which may be verified directly using
(\ref{eq:ss2}) and (\ref{eq:ss1}).

\begin{lemma}\label{lemma:useful}
\begin{rm}
\
\begin{enumerate}
\item  For all $i \ge 1$, $0 < f_i < 1$ and $f_i > f_{i+1}$.
\item $\sum_{i=1}^{\infty} f_i = p$ and
$\sum_{i=1}^{\infty} i f_i = 1$. \quad $\Box$
\end{enumerate}
\end{rm}
\end{lemma}
\medskip

\begin{theorem}\label{theorem:converge}
\begin{rm}
For all $i \ge 1$,
\begin{displaymath}
\lim_{k \to \infty} \frac{E(F_i (k))}{k} = f_i.
\end{displaymath}
\end{rm}
\end{theorem}
{\em Proof.} Using (\ref{eq:epsilon}) to rewrite (\ref{eq:ss0}) and
(\ref{eq:initial}), we obtain, for $i > 1$,
\begin{equation}\label{eqA}
(k+1) (f_i + \epsilon_{i,k+1}) = k (f_i + \epsilon_{i,k}) + k \beta_k
(i - 1 + \alpha) (f_{i-1} + \epsilon_{i-1,k}) - k \beta_k (i +
\alpha) (f_i + \epsilon_{i,k})
\end{equation}
and, for $i = 1$,
\begin{equation}\label{eqB}
(k+1) (f_1 + \epsilon_{1,k+1}) = k (f_1 + \epsilon_{1,k}) - k \beta_k
(1 + \alpha) (f_1 + \epsilon_{1,k}) + p.
\end{equation}
\smallskip

Equations (\ref{eq:ss2}) and (\ref{eq:ss1}) may be written in a
similar form as
\begin{equation}\label{eqC}
(k+1) f_i = k f_i + \beta (i - 1 + \alpha) f_{i-1} - \beta (i +
\alpha) f_i
\end{equation}
and
\begin{equation}\label{eqD}
(k+1) f_1 = k f_1 - \beta (1 + \alpha) f_1 + p.
\end{equation}
\smallskip

For $i > 1$, subtracting (\ref{eqC}) from (\ref{eqA}) yields:
\begin{displaymath}
(k+1) \epsilon_{i,k+1} = k \epsilon_{i,k} + k \beta_k (i - 1 +
\alpha) \epsilon_{i-1,k} - k \beta_k (i + \alpha) \epsilon_{i,k} + (k
\beta_k - \beta) \Big( (i - 1 + \alpha) f_{i-1} - (i + \alpha) f_i
\Big).
\end{displaymath}
\smallskip

Using (\ref{eq:ss2}) and (\ref{eqZ}) this simplifies to
\begin{equation}\label{eqE}
(k+1) \epsilon_{i,k+1} = \Big( 1 - \beta_k (i + \alpha) \Big) k
\epsilon_{i,k} + \beta_k (i - 1 + \alpha) k \epsilon_{i-1,k} - \alpha
\beta_k f_i.
\end{equation}
\smallskip

Similarly, for $i = 1$, subtracting (\ref{eqD}) from (\ref{eqB}) and
using (\ref{eqZ}), we obtain:
\begin{equation}\label{eqF}
(k+1) \epsilon_{1,k+1} = \Big( 1 - \beta_k (1 + \alpha) \Big) k
\epsilon_{1,k} + \alpha \beta \beta_k (1 + \alpha) f_1.
\end{equation}
\smallskip

From (\ref{eqE}), by virtue of (\ref{eqY2}) and the fact that $f_i <
1$ , we have for, $1 < i \le k$,
\begin{equation}\label{eqG}
(k+1) \mid \! \epsilon_{i,k+1} \! \mid \le \Big( 1 - \beta_k (i +
\alpha) \Big) k \mid \! \epsilon_{i,k} \! \mid + \beta_k (i - 1 +
\alpha) k \mid \! \epsilon_{i-1,k} \! \mid + \mid \! \alpha \! \mid
\beta_k.
\end{equation}
\smallskip

From (\ref{eq:epsilon}) it follows that $\epsilon_{i,k} = - f_i$ for
$i > k$, so for $i = k+1$ equation (\ref{eqE}) becomes
\begin{equation}\label{eqKplus1}
(k+1) \epsilon_{k+1,k+1} = \beta_k (k + \alpha) k \epsilon_{k,k} -
f_{k+1} \Big(k \big( 1 - \beta_k (k + \alpha) \big) - k \beta_k +
\alpha \beta_k \Big).
\end{equation}
\smallskip

It follows that

\begin{eqnarray*}
(k+1) \mid \! \epsilon_{k+1,k+1} \! \mid & \le & \beta_k (k + \alpha)
k \mid \! \epsilon_{k,k} \! \mid + f_{k+1} \Big( k \big( 1 - \beta_k
(k + \alpha) \big) + k \beta_k +  \mid \! \alpha \! \mid \beta_k
\Big).
\end{eqnarray*}
\smallskip

Using (\ref{eqZ}) to substitute for $\beta_k$, this gives
\begin{eqnarray}\label{eq:kplus1}
(k+1) \mid \! \epsilon_{k+1,k+1} \! \mid & \le & \frac{\beta (k +
\alpha) k \mid \! \epsilon_{k,k} \! \mid}{k + \alpha \beta} +
\frac{f_{k+1}}{k + \alpha \beta} \Big( k^2 (1 -
\beta) + k \beta + \mid \! \alpha \! \mid \beta \Big) \nonumber \\
&\le & \frac{\beta (k + \alpha) k \mid \! \epsilon_{k,k} \! \mid}{k +
\alpha \beta} + \frac{k}{k + \alpha \beta} \Big(1 + \mid \! \alpha \!
\mid \Big),
\end{eqnarray} since $k \ge 1$, $\beta < 1$, and $k
f_{k+1} < 1$ by Lemma~\ref{lemma:useful}(II). We now define
\begin{equation}\label{eqH}
\delta_k = \max_{i \ge 1} \mid \! \epsilon_{i,k} \! \mid \ = \max_{1
\le i \le k+1} \mid \! \epsilon_{i,k} \! \mid.
\end{equation}
\smallskip
(The two maxima are equal since $\epsilon_{i,k} = - f_i$ for $i > k$,
and $f_i$ is monotonic decreasing.)

\medskip

On using (\ref{eqH}), inequality (\ref{eqG}) yields
\begin{equation}\label{eqJ}
(k+1) \mid \! \epsilon_{i,k+1} \! \mid \le (1 - \beta_k) k \delta_k +
\mid \! \alpha \! \mid \beta_k
\end{equation}
for $1 < i \le k$.

\medskip

Similarly, from (\ref{eqF}), on using (\ref{eqH}) together with
(\ref{eqY1}), (\ref{eqY2}) and Lemma~\ref{lemma:useful}(I), it
follows that
\begin{equation}\label{eqL}
(k+1) \mid \! \epsilon_{1,k+1} \! \mid \le \bigg( 1 - \beta_k (1 +
\alpha)  \bigg) k \delta_k + \mid \! \alpha \! \mid \beta_k (1 +
\alpha).
\end{equation}
\medskip

Now let
\begin{displaymath}
\gamma = \frac{1 + \mid \! \alpha \! \mid}{1 - \beta}
\end{displaymath}
so again in a similar fashion from (\ref{eq:kplus1})
\begin{eqnarray}\label{eq:delta}
(k+1) \mid \! \epsilon_{k+1,k+1} \! \mid & \le & \frac{\beta (k +
\alpha) k \delta_k}{k + \alpha \beta} + \frac{k \gamma (1 - \beta)}{k
+ \alpha \beta}.
\end{eqnarray}
\medskip

We show by induction on $k$ that
\begin{equation}\label{eqK}
k \delta_k \le \gamma.
\end{equation}
\smallskip

From (\ref{eq:epsilon}) and (\ref{eqH}) we see that $\delta_1 = \max
\{1 - f_1, f_2\}$. So, by Lemma~\ref{lemma:useful}(I), (\ref{eqK})
holds for $k=1$.

\medskip

Now assume that (\ref{eqK}) holds for some $k \ge 1$. So, for $1 < i
\le k$, since $\mid \! \alpha \! \mid < \gamma$, (\ref{eqJ}) gives
\begin{equation}\label{eqM}
(k+1) \mid \! \epsilon_{i,k+1} \! \mid \le (1 - \beta_k) \gamma +
\mid \! \alpha \! \mid \beta_k \le \gamma.
\end{equation}
\medskip

For $i = k+1$ from (\ref{eq:delta}) using (\ref{eqH}) and
(\ref{eqK}), we have
\begin{eqnarray}\label{eq:fin1}
(k+1) \mid \! \epsilon_{k+1,k+1} \! \mid & \le & \frac{\beta (k +
\alpha) \gamma}{k + \alpha \beta} + \frac{k \gamma (1 - \beta)}{k +
\alpha \beta} = \gamma.
\end{eqnarray}
\medskip

For $i > k+1$, since $\epsilon_{i,k+1} = - f_i$, by
Lemma~\ref{lemma:useful} (II),
\begin{equation}\label{eq:fin2}
(k+1) \mid \! \epsilon_{i,k+1} \! \mid = (k+1) f_i < 1 \le \gamma.
\end{equation}
\smallskip

Similarly, for $i = 1$, (\ref{eqL}) gives
\begin{equation}\label{eqN}
(k+1) \mid \! \epsilon_{1,k+1} \! \mid \le \bigg( 1 - \beta_k (1 +
\alpha) \bigg) \gamma + \mid \! \alpha \! \mid \beta_k (1 + \alpha)
\le \gamma.
\end{equation}
\smallskip

Therefore, from (\ref{eqM}), (\ref{eq:fin1}), (\ref{eq:fin2}) and
(\ref{eqN}), $(k+1) \delta_{k+1} \le \gamma$.

\medskip

So, by induction, (\ref{eqK}) holds for all $k \ge 1$. Thus, to
conclude the proof, we note that, as $k$ tends to infinity,
$\delta_k$ tends to $0$, so $\epsilon_{i,k}$ tends to $0$ for all
$i$. \quad $\Box$

\medskip

\newcommand{\etalchar}[1]{$^{#1}$}

\end{document}